\documentclass[%
 aip,
pof,
 amsmath,amssymb,
preprint,%
]{revtex4-1}

\usepackage {CJK}

\usepackage{graphicx}% Include figure files
\usepackage{dcolumn}% Align table columns on decimal point
\usepackage{bm}% bold math
\usepackage[utf8]{inputenc}
\usepackage[T1]{fontenc}
\usepackage{mathptmx}

\usepackage{url}

\begin{document}

\begin{CJK*}{UTF8}{} % Use default fonts from CJK (see below)

\preprint{Preprint submit to \emph{Physics of Fluids}}

\title{Transport and deposition of dilute microparticles in turbulent thermal convection}
% Force line breaks with \\
%%\email{}

\author{Ao Xu}
 \email{axu@nwpu.edu.cn}
 \affiliation{School of Aeronautics, Northwestern Polytechnical University, Xi'an 710072, China}%

\author{Shi Tao}
 \affiliation{Key Laboratory of Distributed Energy Systems of Guangdong Province, Dongguan University of Technology, Dongguan 523808, China}%

\author{Le Shi}
 \affiliation{State Key Laboratory of Electrical Insulation and Power Equipment, Center of Nanomaterials for Renewable Energy, School of Electrical Engineering, Xi'an Jiaotong University, Xi'an 710049, China}%

\author{Heng-Dong Xi}
 \affiliation{School of Aeronautics, Northwestern Polytechnical University, Xi'an 710072, China}%

\date{\today}% It is always \today, today,
             %  but any date may be explicitly specified

\begin{abstract}
We analyze the transport and deposition behavior of dilute microparticles in turbulent Rayleigh-B\'enard convection.
Two-dimensional direct numerical simulations were carried out for the Rayleigh number ($Ra$) of $10^{8}$ and the Prandtl number ($Pr$) of 0.71 (corresponding to the working fluids of air).
The Lagrangian point particle model was used to describe the motion of microparticles in the turbulence.
Our results show that the suspended particles are homogeneously distributed in the turbulence for the Stokes number ($St$) less than $10^{-3}$, and they tend to cluster into bands for $10^{-3} \lesssim St \lesssim 10^{-2}$.
At even larger $St$, the microparticles will quickly sediment in the convection.
We also calculate the mean-square displacement (MSD) of the particle's trajectories.
At short time intervals, the MSD exhibits a ballistic regime, and it is isotropic in vertical and lateral directions;
at longer time intervals, the MSD reflects a confined motion for the particles, and it is anisotropic in different directions.
We further obtained a phase diagram of the particle deposition positions on the wall, and we identified three deposition states depending on the particle's density and diameter.
An interesting finding is that the dispersed particles preferred to deposit on the vertical wall where the hot plumes arise, which is verified by tilting the cell and altering the rotation direction of the large-scale circulation.
\footnote{This article may be downloaded for personal use only.
Any other use requires prior permission of the author and AIP Publishing.
This article appeared in Xu et al., Phys. Fluids \textbf{32}, 083301 (2020) and may be found at \url{https://doi.org/10.1063/5.0018804}.}
\end{abstract}

\maketitle
\end{CJK*}

\section{\label{sec:level1}Introduction}

Transport and deposition of solid particles (or liquid droplets) in turbulent thermal convection occur ubiquitously in environmental science \cite{guha2008transport,toschi2009lagrangian,tenneti2014particle,mathai2020bubbly}.
For example, suspended atmospheric pollutant particles (PM10 and PM2.5) that originated from dust and smoke will severely influence the air quality \cite{seinfeld2016atmospheric,norback2019sources}.
Another example is pathogen laden droplets in confined indoors, which will cause viral and bacterial infectious diseases (SARS and  COVID-19) to spread in hospitals, schools, and airplanes  \cite{bourouiba2014violent,mittal2020flow,dbouk2020coughing,chaudhuri2020modeling}.
In such a dispersed multiphase flow, the evolution of the phase interface may not be a primary concern \cite{balachandar2010turbulent}.
From the aspect of particle kinematics, important control parameters include the density ratio of the particle to its surrounding fluid, $\Gamma=\rho_{p}/\rho_{f}$, and the size ratio $\Xi=d_{p}/l_{f}$.
Here, $\rho_{p}$ and $d_{p}$ are the particle density and particle size, respectively.
$\rho_{f}$ is the fluid density, and $l_{f}$ is the characteristic fluid length.
When $\Xi \ll 1$, the Lagrangian particle model can be used to track the dispersed phase.
Moreover, when the volume fraction of the dispersed phase is small, the dominant effect is that of the carrier flow on the dynamics of the dispersed phase, but not vice versa.
Thus, a one-way interphase coupling approach can be adopted to track the motions of particles \cite{van2008numerical,maxey2017simulation}.
Previous studies have shown that even in homogeneous isotropic turbulence, the dispersed particles may not distribute homogeneously but exhibit preferential concentration \cite{wang1993settling,bosse2006small,calzavarini2008dimensionality,zhang2016preferential}.
For light particles with a density ratio of $\Gamma \ll 1$, they concentrate in regions of high vorticity; for heavy particles with a density ratio of $\Gamma \gg 1$, they are expelled from rotating regions.

Due to the injected buoyancy and the effect of the domain boundaries, turbulent thermal convection is generally inhomogeneous and anisotropic.
A simple paradigm system to study thermal convection is the Rayleigh-B\'enard (RB) cell, where a fluid layer is heated from the bottom and cooled from the top \cite{ahlers2009heat,lohse2010small,chilla2012new,xia2013current,mazzino2017two,wang2020vibration,kumar2018physics}.
The control parameters of the RB system include the Rayleigh number $Ra = \beta g \Delta_{T}H^{3}/(\nu_{f}\kappa_{f})$ and the Prandtl number $Pr = \nu_{f}/\kappa_{f}$.
The $Ra$ describes the strength of buoyancy relative to thermal and viscous dissipative effects.
The $Pr$ describes thermophysical fluid properties.
Here, $\beta$, $\kappa_{f}$, and $\nu_{f}$ are the thermal expansion coefficient, thermal diffusivity, and kinematic viscosity of the fluid, respectively.
$g$ is the gravitational acceleration.
$\Delta_{T}$ is the imposed temperature difference between the top and bottom fluid layers of height $H$.
In the RB convection, ubiquitous coherent structures include thermal plumes and large-scale circulation (LSC) \cite{hiroaki1980turbulent,krishnamurti1981large}.
Specifically, sheet-like plumes that detached from boundary layers transform into mushroom-like ones via mixing, merging, and clustering \cite{zhou2007morphological}.
Due to plume-vortex and plume-plume interactions, thermal plumes further self-organize into the LSC that spans the size of the convection cell \cite{xi2004laminar}.

Although the dynamics of single-phase turbulent thermal convection has been thoroughly investigated, the complex interactions between dispersed immiscible phase and its surrounding fluid in turbulent thermal convection remain less explored.
One of the few studies by Puragliesi et al. \cite{puragliesi2011dns} focused on particle deposition in side-heated convection cells (i.e., heated from one vertical side and cooled from the other vertical side).
They found that a strong recirculating zone contributes to the decreased gravitational settling, thus resulting in particles suspending with a longer time.
Because the driven force, namely, the temperature gradient, in the side-heated convection cell is perpendicular to that in the RB convection cell, the fluid and particle dynamics are expected to be different in these two cells.
Lappa \cite{lappa2018transport} analyzed the pattern produced by inertial particles dispersed in the localized rising thermal plume.
He identified the average behavior of particles by revealing the mean evolution.
It should be noted that although the thermal plumes are the building blocks of turbulent thermal convection, the LSC, which is another essential feature of the turbulent thermal convection, is missing in such analysis.
In addition to the one-way coupling between the dispersed phase and the carrier flow, Park et al. \cite{park2018rayleigh} further investigated the RB turbulence modified by inertial and thermal particles.
Changes of the integrated turbulent kinetic energy and heat transfer efficiency were quantified.
The results showed that particles with the Stokes number (to be defined in Sec. \ref{subsectionSettings}) of order unity maximize the heat transfer efficiency.
However, particles with such a high Stokes number (either heavy density or large size) will sediment quickly in the air, which may be of limited interest for studying suspended atmospheric pollutant particles or pathogen laden droplets.

In this work, our objective is to shed light on the dynamics of atmospheric pollutant particles or pathogen laden droplets.
We simulate transport and deposition of dilute microparticles in an RB convection cell with air as the working fluid (i.e., $Pr=0.71$) at a high $Ra$ number (i.e., $Ra=10^{8}$) such that the ubiquitous features of the turbulent thermal convection (including thermal plumes and the LSC) naturally arise.
We choose the typical particle parameters as 10 $\mu$m $\le$ $d_{p}$ $\le$ 100 $\mu$m and 400 kg/m$^{3}$ $\le \rho_{p} \le$ 4000 kg/m$^{3}$, and the corresponding particle Stokes number (i.e., $3.67\times10^{-4} \le St \le 0.37$) is much lower than that by Park et al. \cite{park2018rayleigh} (i.e., $0.1 < St < 15$).
The rest of this paper is organized as follows:
In Sec. \ref{Section2}, we present the numerical details for the simulations, including the direct numerical simulation of thermal turbulence and the Lagrangian point particle model.
In Sec. \ref{Section3}, we analyzed the particle transport behavior via flow visualization and particle mean-square displacement calculation, followed by the statistics of particle deposition behavior, such as the time history of the particle deposition ratio and the phase diagram of the particle deposition location.
In Sec. \ref{SectionConclusions}, the main findings of the present work are summarized.

\section{\label{Section2}Numerical method}

\subsection{Numerical model for incompressible thermal flows}

In incompressible thermal flows, temperature variation will cause density variation, thus resulting in a buoyancy effect.
Following the Boussinesq approximation, the temperature can be treated as an active scalar, and its influence on the velocity field is realized through the buoyancy term.
The governing equations can be written as
\begin{subequations}
\begin{align}
& \nabla \cdot \mathbf{u}_{f}=0 \\
& \frac{\partial \mathbf{u}_{f}}{\partial t}+\mathbf{u}_{f}\cdot \nabla \mathbf{u}_{f}=-\frac{1}{\rho_{0}}\nabla p+\nu_{f} \nabla^{2}\mathbf{u}_{f}+g\beta(T-T_{0})\hat{\mathbf{y}} \\
& \frac{\partial T}{\partial t}+\mathbf{u}_{f}\cdot \nabla T=\kappa \nabla^{2} T
\end{align} \label{Eq.NS}
\end{subequations}
where $\mathbf{u}_{f}$, $p$ and $T$ are the fluid velocity, pressure and temperature, respectively.
$\rho_{0}$ and $T_{0}$  are the reference density and temperature, respectively.
$\hat{\mathbf{y}}$  is the unit vector in the vertical direction.
In the above equations, all the transport coefficients are assumed to be constants.

We adopt the lattice Boltzmann (LB) method \cite{chen1998lattice,aidun2010lattice,xu2017lattice,huang2015multiphase} as the numerical tool for the direct numerical simulation of turbulent thermal convection.
The advantages of the LB method include easy implementation and parallelization as well as low numerical dissipation \cite{xu2017accelerated}.
In the LB method, to solve Eqs. 1a and 1b, the evolution equation of the density distribution function is written as \cite{chen1998lattice,aidun2010lattice}
\begin{equation}
  f_{i}(\mathbf{x}+\mathbf{e}_{i}\delta_{t},t+\delta_{t})-f_{i}(\mathbf{x},t)=-(\mathbf{M}^{-1}\mathbf{S})_{ij}\left[\mathbf{m}_{j}(\mathbf{x},t)-\mathbf{m}_{j}^{(\text{eq})}(\mathbf{x},t)\right]
  +\delta_{t}F_{i}^{'} \label{Eq.MRT}
\end{equation}
To solve Eq. \ref{Eq.NS}c, the evolution equation of temperature distribution function is written as \cite{chen1998lattice,aidun2010lattice}
\begin{equation}
  g_{i}(\mathbf{x}+\mathbf{e}_{i}\delta_{t},t+\delta_{t})-g_{i}(\mathbf{x},t)=-(\mathbf{N}^{-1}\mathbf{Q})_{ij}\left[\mathbf{n}_{j}(\mathbf{x},t)-\mathbf{n}_{j}^{(\text{eq})}(\mathbf{x},t)\right]
  \label{Eq.MRT_T}
\end{equation}
Here, $f_{i}$ and $g_{i}$ are the density and temperature distribution function, respectively.
$\mathbf{x}$ is the fluid parcel position, $t$ is the time, and $\delta_{t}$ is the time step.
$\mathbf{e}_{i}$ is the discrete velocity along the $i$th direction.
$\mathbf{M}$ is a $9\times 9$ orthogonal transformation matrix based on the D2Q9 discrete velocity model;
$\mathbf{N}$ is a $5\times 5$ orthogonal transformation matrix based on the D2Q5 discrete velocity model.
The equilibrium moments $\mathbf{m}^{(\text{eq})}$ in Eq. 2 are
\begin{equation}
\mathbf{m}^{(\text{eq})}=\rho \left[ 1, \ -2+3|\mathbf{u}_{f}|^{2}, \ 1-3|\mathbf{u}_{f}|^{2}, \ u_{f}, \ -u_{f}, \
v_{f}, \ -v_{f}, \ 2u_{f}^{2}-v_{f}^{2}, \ u_{f}v_{f} \right]^{T}
\end{equation}
The equilibrium moments $\mathbf{n}^{(\text{eq})}$ in Eq. 3 are
\begin{equation}
\mathbf{n}^{(\text{eq})}=\left[ T, \ u_{f}T, \ v_{f}T, \ a_{T}T, 0 \right]^{T}
\end{equation}
where $a_{T}$ is a constant determined by thermal diffusivity as $a_{T}=20\sqrt{3}\kappa-6$.
The relaxation matrix $\mathbf{S}$ is $\mathbf{S}=\text{diag}(s_{\rho},s_{e},s_{\varepsilon},s_{j},s_{q},s_{j},s_{q},s_{\nu},s_{\nu})$,
and the kinematic viscosity of the fluid is calculated as $\nu=c_{s}^{2}(\tau_{f}-0.5)$.
The relaxation matrix $\mathbf{Q}$ is $\mathbf{Q}=\text{diag}(0,q_{\kappa},q_{\kappa},q_{e},q_{\nu})$,
where $q_{\kappa}=3-\sqrt{3}$, $q_{e}=q_{\nu}=4\sqrt{3}-6$.

The macroscopic fluid variables of density $\rho$, velocity $\mathbf{u}_{f}$,  and temperature $T$ are calculated as $\rho=\sum_{i=0}^{8}f_{i}, \ \ \mathbf{u}=\left( \sum_{i=0}^{8}\mathbf{e}_{i}f_{i}+\mathbf{F}/2 \right)/\rho$, and $T=\sum_{i=0}^{4}g_{i}$, respectively.
More numerical details on the LB method and validation of the in-house code can be found in our previous work \cite{xu2018thermal,xu2019lattice,xu2019statistics}.

\subsection{Kinematic equation for the particles}
We consider small particles such that their presences does not modify the turbulence structure, namely, one-way coupling between the multiphase.
Here, 'small' means the diameter of the particle is smaller than the Kolmogorov length scale of the turbulence; however, the diameter of the particle should still be much larger than the molecular mean free path such that the effect of Brownian motion can be neglected.
In addition, the particles are assumed to be isotropic such that we only consider the motion of the particle and neglect the rotation of the particle \cite{voth2017anisotropic,calzavarini2020anisotropic}.
Specifically, the particles' motions are described by Newton's second law as
\begin{equation}
m_{p}\frac{d\mathbf{u}_{p}(t)}{dt}=\mathbf{F}_{\text{total}}(t)=\mathbf{F}_{G}(t)+\mathbf{F}_{D}(t)
\end{equation}

The total force $\mathbf{F}_{\text{total}}$ exerted on the particle includes the net gravitational force $\mathbf{F}_{G}$ and the drag force $\mathbf{F}_{D}$.
Specifically, particles experience a gravitational force in the direction of gravitational acceleration, as well as buoyancy in the opposite direction.
The net gravitational force $\mathbf{F}_{G}$ is given by
\begin{equation}
\mathbf{F}_{G}=\rho_{p}V_{p}\mathbf{g}-\rho_{f}V_{p}\mathbf{g}
\end{equation}
where $\rho_{p}$ and $V_{p}$ are the density and volume of the particle, respectively.
Meanwhile, the particle experiences a drag force that acts to catch up with the changing velocity of the surrounding fluid.
The drag force $\mathbf{F}_{D}$ is given by
\begin{equation}
\mathbf{F}_{D}=\frac{m_{p}}{\tau_{p}}\left(\mathbf{u}_{f}-\mathbf{u}_{p} \right)f(Re_{p})
\end{equation}
where $m_{p}$ and $\mathbf{u}_{p}$ are the mass and velocity of the particle, respectively.
$\tau_{p}=\rho_{p}d_{p}^{2}/(18\mu_{f})$ is the particle response time, and $d_{p}$ is the particle diameter.
The particle Reynolds number $Re_{p}=d_{p}|\mathbf{u}_{f}-\mathbf{u}_{p}|/\nu_{f}$ determines the coefficient $f(Re_{p})$.
When $Re_{p}$ is much less than 1, namely, a Stokes drag law is valid, we have $f(Re_{p}) \approx 1$.
In general, Clift et al. \cite{clift1978bubbles} gave the relationship $f(Re_{p})=1+0.15Re_{p}^{0.687}$  for $Re_{p} < 40$.

\subsection{\label{subsectionSettings}Simulation settings}

\begin{figure}[htbp]
\centering
\includegraphics[width=12cm]{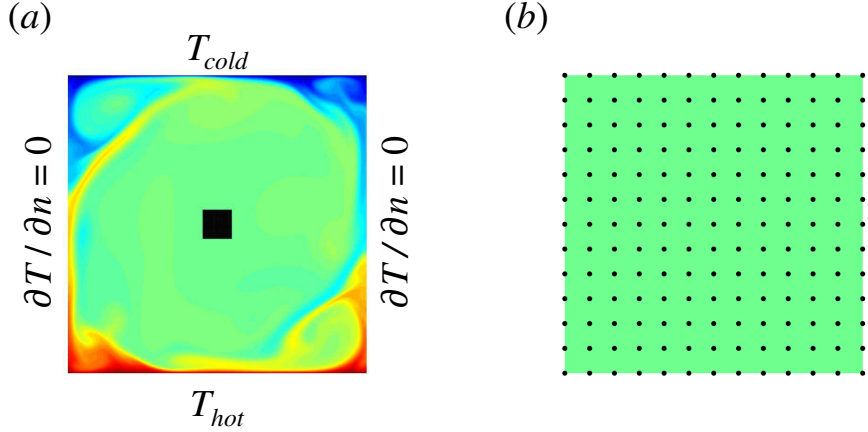}
\caption{\label{initial} (\textit{a}) Illustration of the particles' initial positions in the convection cell, the contour represents the instantaneous temperature field; (\textit{b}) an enlarged view of the central region in (\textit{a}). The black dots represent the particles whose sizes have been artificially increased for the convenience of flow visualization.}
\end{figure}

We consider the particle motions in a 2D convection cell with a size $H\times H$.
The top and bottom walls of the cell are kept at a constant cold and hot temperatures, respectively; the other two vertical walls are adiabatic.
All four walls impose no-slip velocity boundary conditions.
Our simulation protocol is as follows:
We start the simulation of single-phase turbulent thermal convection, namely, without considering the particles' motion.
The particles are released in the turbulence after a statistically stationary state has reached, which takes 500 $t_{f}$.
Here, $t_{f}=\sqrt{H/(g\beta \Delta_{T})}$ denotes free-fall time units.
We then advance the fluid flows and the motion of the particles simultaneously.
A total number of 10 000 particles are initially placed at the cell central region (see Fig. \ref{initial} for the illustration, the 10 000 particles are initially grouped into a $100\times 100$ array, and each particle is placed half grid spacing away from the other).
The initial velocities of the particles are equal to that of the local fluid.
The initial particle configuration approximates the transport of pollutant particles emitted from a source, and the dilute particles may mimic the particle-laden fluid in a cough \cite{duguid1946size}.
We average 2000 $t_{f}$  to obtain statistics for the turbulent flows and the particles.
When a particle hits the wall, we assume that it will deposit on the wall and no longer transport in the convection cell.

We provide the simulation results for a fixed Rayleigh number of $Ra = 10^{8}$ and a Prandtl number of $Pr = 0.71$ (corresponding to the working fluids of air at 300 K).
The other detailed simulation parameters are listed in Table 1.
The mesh size is $513\times 513$ such that the grid spacing $\Delta_{g}$ and time interval $\Delta_{t}$ are properly resolved to compare with the Kolmogorov and Batchelor scales.
Here, the Kolmogorov length scale is estimated by the global criterion $\eta_{K}=\left(\nu^{3}/\langle \varepsilon_{u} \rangle_{V,t}\right)^{1/4}=HPr^{1/2}/\left[Ra(Nu-1)\right]^{1/4}$, the Batchelor length scale is estimated by $\eta_{B}=\eta_{K} Pr^{-1/2}$, and the Kolmogorov time scale is estimated by $\tau_{\eta_{K}}=\sqrt{\nu/\langle \varepsilon_{u} \rangle_{V,t}}=t_{f}\sqrt{Pr/(Nu-1)}$.
The global heat transport is measured by the volume-averaged Nusselt number as $Nu=1+\sqrt{Pr Ra}\langle vt \rangle_{V,t}$, while the Reynolds number $Re=\sqrt{\langle u^{2}+v^{2}\rangle_{V,t}}H/\nu$ measures the global strength of the convection.
Here, $\langle \cdots \rangle_{V,t}$ denotes the volume and time average.
$\varepsilon_{u}$ denotes the kinetic energy dissipation rates, and its global average can be related to the Nusselt number via \cite{shraiman1990heat} the exact relation  $\langle \varepsilon_{u} \rangle_{V,t}=\nu^{3}Ra(Nu-1)/(H^{4}Pr^{2})$.
The simulation results have shown that grid spacing satisfies $\max(\Delta_{g}/\eta_{K}, \Delta_{g}/\eta_{B}) \le 0.51$, which ensures the spatial resolution; the time intervals are  $\Delta_{t}\le 0.0006\tau_{\eta_{K}}$, thus adequate temporal resolution is guaranteed.
In addition, our results for Nusselt and Reynolds numbers (i.e., $Nu = 25.36$, $Re = 3602$) are consistent with the previous results reported by Zhang et al. \cite{zhang2017statisticsJFM} (i.e., $Nu = 25.25$, $Re =3662$).

\begin{table}[htbp]
\caption{Fluid properties and simulation parameters.}
\begin{tabular}{cc}
  \hline
  Parameter & Value \\
  \hline
  Rayleigh number ($Ra$) & $10^{8}$ \\
  Prandtl number ($Pr$)  & 0.71 \\
  Reference temperature ($T_{0}$) & 300 K \\
  Reference fluid density ($\rho_{0}$) & 1.18 kg/m$^{3}$ \\
  Thermal expansion coefficient ($\beta$) & $3.36\times 10^{-3}$ K$^{-1}$ \\
  Kinematic viscosity ($\nu_{f}$)  & $1.58\times 10^{-5}$ m$^{2}$/s \\
  Thermal diffusivity ($\kappa_{f}$) & $2.21 \times 10^{-5}$ m$^{2}$/s \\
  Temperature differences ($\Delta_{T}$) & 5 K \\
  Cell size ($H$)  & 0.60 m\\
  \hline
\end{tabular}
\end{table}

\begin{figure}[htbp]
\centering
\includegraphics[width=11cm]{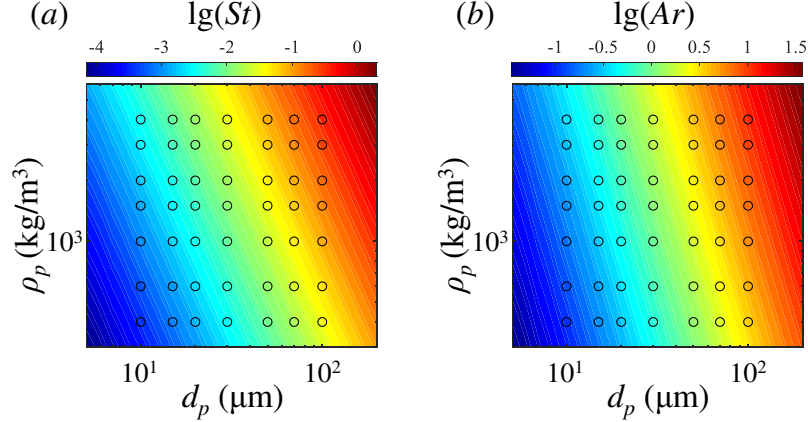}
\caption{\label{paraSpace} (\textit{a}) The logarithmic of the particle Stokes number and (\textit{b}) the logarithmic of the Archimedes number as functions of the particle diameter $d_{p}$ and particle density $\rho_{p}$. The black circles represent our simulation parameters.}
\end{figure}

In the simulations, the non-dimensional control parameters for the particles include the density ratio of the particle to its surrounding fluid $\Gamma=\rho_{p}/\rho_{f}$ and the size ratio $\Xi=d_{p}/l_{f}$.
By combing the $\Gamma$ and $\Xi$, we can obtain the particle Stokes number ($St$) and the Archimedes number ($Ar$) as
\begin{equation}
St=\frac{\tau_{p}}{\tau_{\eta}}=\frac{\rho_{p}d_{p}^{2}/(18\mu_{f})}{\sqrt{\nu/\langle \varepsilon_{u} \rangle_{V,t}}}, \ \ \ \ \
Ar=\sqrt{\frac{\rho_{p}-\rho_{f}}{\rho_{f}}\frac{gd_{p}^{3}}{\nu^{2}}}
\end{equation}
where $\tau_{p}$ is the particle response time.
The $St$ describes the particle inertia relative to that of the fluid, and the $Ar$ describes the ratio of gravity forces to the viscous forces.
Because we have fixed the $Ra$ and the $Pr$ in the simulation, namely, thermal convection related quantities are fixed, we then have $St \propto \rho_{p}$, $Ar \propto \rho_{p}^{1/2}$ and  $St \propto d_{p}^{2}$, $Ar \propto d_{p}^{3/2}$.
The $St$ and the $Ar$ numbers can be uniquely determined by $d_{p}$ and $\rho_{p}$, as shown in Fig. \ref{paraSpace}.
We explore the parameter space of 10 $\mu$m $\le d_{p} \le 100$ $\mu$m and 400 kg/m$^{3}$ $\le \rho_{p} \le$ 4000 kg/m$^{3}$, denoted by the black circles in Fig. \ref{paraSpace}.
We note that the estimated Kolmogorov length scale is $\eta_{K}=2.27$ mm, and the largest particle volume fraction of all cases is only 0.02$\%$.
Thus, for dilute particles with diameters fall in the range mentioned above, the one-way coupling strategy is justified to model their motions.
For particles with larger numbers but still similar size, i.e., particles with higher particle volume fraction, a four-way coupling strategy is necessary to describe the interactions between the particle and its surrounding fluid \cite{akiki2017pairwiseJFM,akiki2017pairwiseJCP}.

\section{\label{Section3}Results and discussion}

\subsection{\label{Section31}Particle transport in the convection cell}

\begin{figure}[htbp]
\centering
\includegraphics[width=13cm]{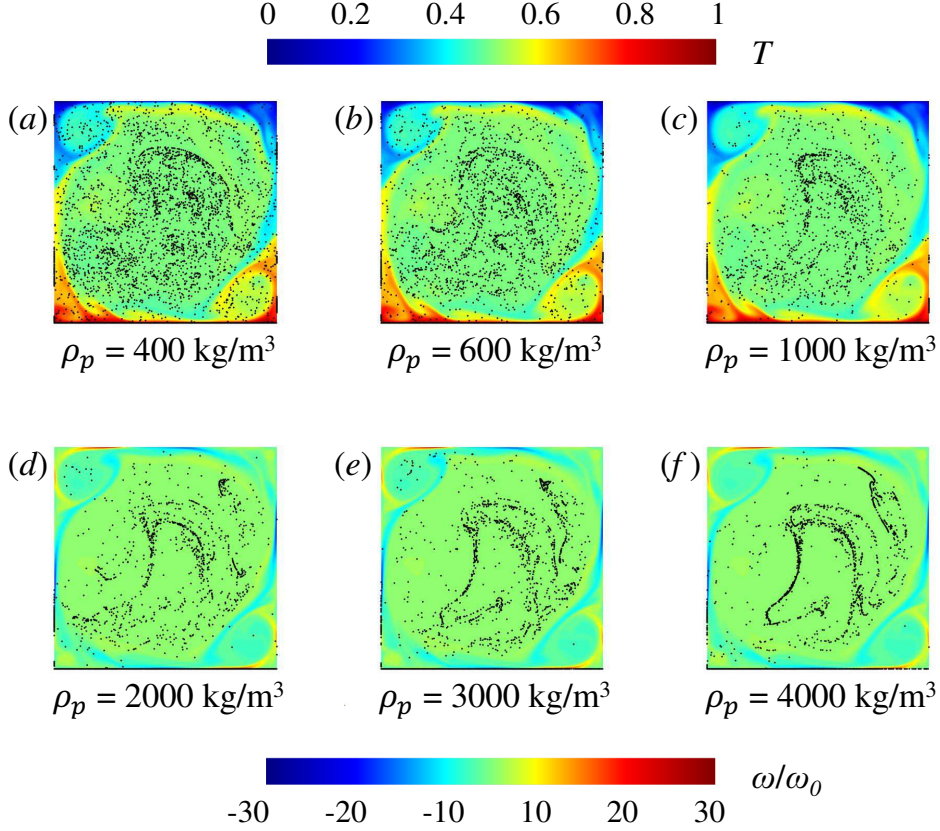}
\caption{\label{dp10um} Snapshots of the particles' positions at $t = 500 \ t_{f}$. Panels (\textit{a})-(\textit{c}) show the temperature field (contour), while panels (\textit{d})-(\textit{f}) show the vorticity field (contour). $\omega_{0}$ denotes the instantaneous vorticity at the cell center. The diameters of these particles are 10 $\mu$m.
%%and their densities are ($\textit{a}$) 400 kg/m$^{3}$, ($\textit{b}$) 600 kg/m$^{3}$, ($\textit{c}$) 1,000 kg/m$^{3}$, ($\textit{d}$) 2,000 kg/m$^{3}$, ($\textit{e}$) 3,000 kg/m$^{3}$, ($\textit{f}$) 4,000 kg/m$^{3}$.
}
\end{figure}

Figure \ref{dp10um} shows the snapshots of the instantaneous particles' positions for $d_{p} = 10$ $\mu$m and 400 kg/m$^{3}$ $\le \rho_{p} \le$ 4000 kg/m$^{3}$ (corresponding to $3.67\times 10^{-4} \le St \le 3.67\times 10^{-3}$ and $0.115 \le Ar \le 0.366$) at $t = 500 \ t_{f}$ (corresponding to $t \approx 951.6$ s).
Here, we denote the time origin $t = 0$ as the instant when the particles are released in the turbulence.
At such small $St$ and $Ar$, the particles' motions are profoundly affected by the LSC of the convection.
Specifically, these relatively small particles are well dispersed in the turbulence, and they can remain suspended for a long time.
On the other hand, we also notice the differences in the spatial pattern of particles' positions:
the particles are more homogeneously distributed in the turbulence at relatively smaller particle density [see Figs. \ref{dp10um}(a)-\ref{dp10um}(c), which corresponds to $3.67\times 10^{-4} \le St \le 9.19\times 10^{-4}$ and $0.115 \le Ar \le 0.183$].
In contrast, they tend to cluster into bands at relatively larger particle density [see Figs. \ref{dp10um}(d)-\ref{dp10um}(f), which corresponds to $1.84\times 10^{-3} \le St \le 3.67\times 10^{-3}$ and $0.259 \le Ar \le 0.366$].
The clustered particle are repelled from regions of high vorticity, as visualized by the contour of vorticity $\omega=\nabla \times \mathbf{u}$ in Figs. \ref{dp10um}(d)-\ref{dp10um}(f), which shows similar pattern (but at much smaller $St$) compared to those in homogeneous isotropic turbulence \cite{wang1993settling,bosse2006small,calzavarini2008dimensionality,zhang2016preferential}.
We also notice that there are fewer particles in the corner rolls of the convection with the increase in particle density.
The previous study by Park et al. \cite{park2018rayleigh} indicates that the clustering behavior in thermal turbulence occurs at much larger particle $St$ number (namely, $St \approx 1$) when the dimensionless particle settling velocity $V_{g}/U_{buoy}=\left[ \rho_{p}d_{p}^{2}g/(18\mu_{f}) \right]/\sqrt{g\beta \Delta H}$ is fixed as 0.001.
However, if we assume the carrier fluid is air, a quantitative estimation shows that simultaneously achieving $St \approx 1$ and $V_{g}/U_{buoy}=0.001$ would result in an artificially tiny gravity value (almost eight orders of magnitude smaller than 9.8 m/s$^{2}$).

\begin{figure}[htbp]
\centering
\includegraphics[width=13cm]{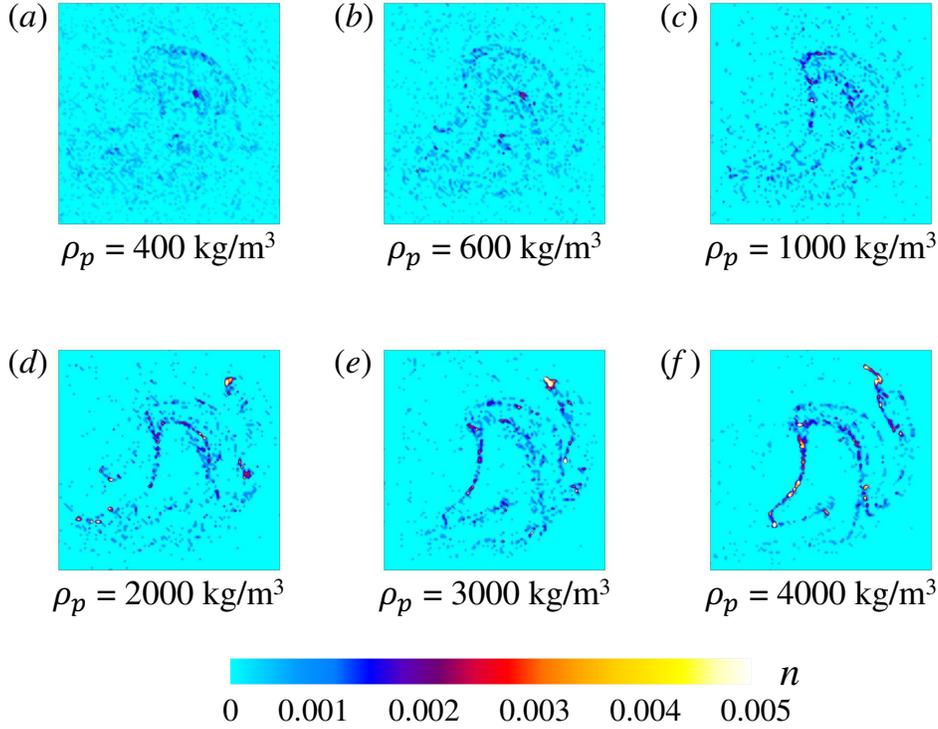}
\caption{\label{dp10um_particleDensity} Snapshots of the instantaneous local particle number density at $t = 500 \ t_{f}$. The diameters of these particles are 10 $\mu$m, and their densities are ($\textit{a}$) 400 kg/m$^{3}$, ($\textit{b}$) 600 kg/m$^{3}$, ($\textit{c}$) 1000 kg/m$^{3}$, ($\textit{d}$) 2000 kg/m$^{3}$, ($\textit{e}$) 3000 kg/m$^{3}$, and ($\textit{f}$) 4000 kg/m$^{3}$.
}
\end{figure}

The above-mentioned flow visualizations illustrate the preferential distribution of particles in the thermal turbulence.
To quantitatively describes the spatial distribution of the particles, we divide the simulation domain into $100 \times 100$ uniform subcells and calculate the local particle number density as
\begin{equation}
n(i,j,t)=\frac{N(i,j,t)}{N_{\text{total}}(t)}
\end{equation}
where $N(i,j,t)$ is the number of suspended particles found inside the ($i,j$)th small square subcell (here $1 \le i,j \le 100$)
and $N_{\text{total}}(t)$ is the number of suspended particle in the whole convection cell at time $t$.
In Fig. \ref{dp10um_particleDensity}, we plot the local particle number density at $t=500 \ t_{f}$, where we can observe homogenous local particle number densities for 400 kg/m$^{3} \le \rho_{p} \le$ 1000 kg/m$^{3}$.
The local particle number densities are more inhomogeneous for 2000 kg/m$^{3} \le \rho_{p} \le$ 4000 kg/m$^{3}$, which is due to higher particle inertia and longer particle response time to the carrier flow.

\begin{figure}[htbp]
\centering
\includegraphics[width=9cm]{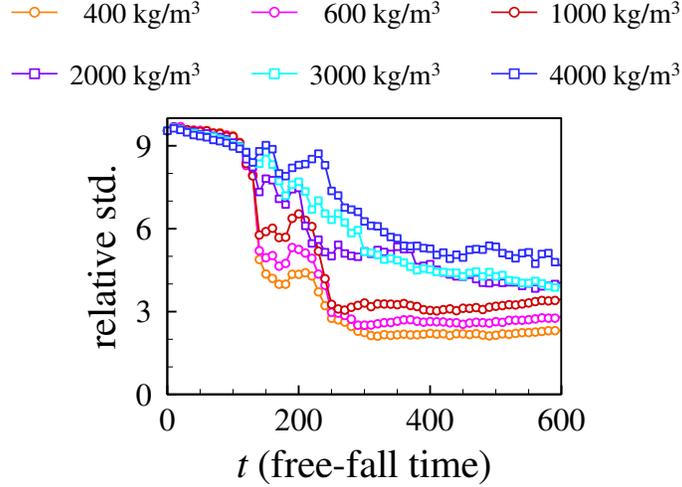}
\caption{\label{varCoeff} Time histories of the relative standard deviation for local particle number density (particles with $d_{p}=10\ \mu$m).}
\end{figure}

We further calculate the relative standard deviation of the local particle number density, namely, the root-mean-square (r.m.s.) of particle number density normalized by the volume-averaged particle number density, which is defined as
\begin{equation}
\text{relative std.} = \frac{1}{\bar{n}(t)}\sqrt{\frac{\sum_{i,j}\left[ n(i,j,t)-\bar{n}(t) \right]^{2}}{100\times 100}}
\end{equation}
Here, $\bar{n}(t)$ denotes the volume-averaged particle number density at time $t$.
In Fig. \ref{varCoeff}, we plot the time histories of the relative standard deviation for particles with a diameter of 10 $\mu$m.
We can see that the deviations decrease rapidly during the initial transient state (i.e., $t \lesssim 250 \ t_{f}$), which is due to the dispersion of the particle group after being released in the turbulence.
At $t \gtrsim 250 \ t_{f}$, the relative standard deviations nearly reach a plateau, indicating the good dispersion of the particles in the turbulence.
We also found that the relative standard deviation of the local particle number density depends on the $St$ and $Ar$, as light density and small size of the particles favor their dispersion.

\begin{figure}[htbp]
\centering
\includegraphics[width=13cm]{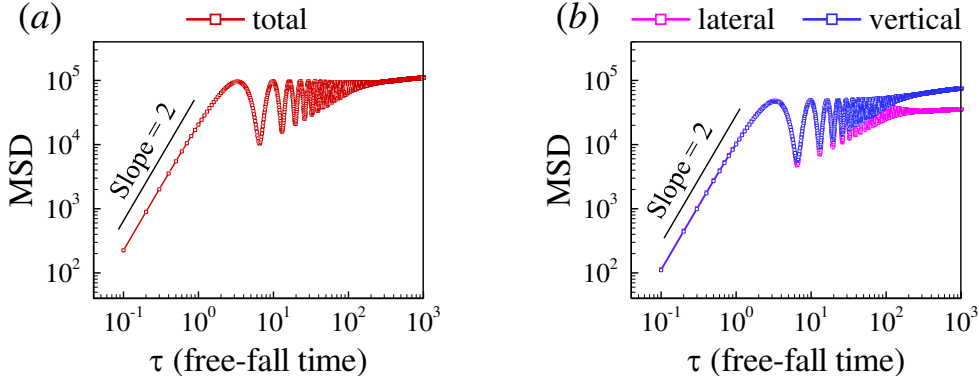}
\caption{\label{MSD} (\textit{a}) The total mean-square displacement (MSD) of particles' trajectories and (\textit{b}) the MSD in the lateral and vertical directions for particles with $d_{p}$ = 10 $\mu$m and $\rho_{p}$ = 1000 kg/m$^{3}$.}
\end{figure}

We then analyze the statistics of particles' trajectories by calculating their mean-square displacement $\text{MSD}(\tau)=\langle [\mathbf{r}(t+\tau)-\mathbf{r}(t)]^{2} \rangle$.
Here, $\mathbf{r}(t)$ is the particle's position at time $t$ and $\tau$ is the lag time between the two positions taken by the particles.
The average $\langle \cdots \rangle$ represents a time-average over $t$ and an ensemble-average over trajectories.
When a particle is deposited on the wall, we will stop tracking its trajectory.
Figure \ref{MSD}(a) shows the MSD for particles with $d_{p} = 10$ $\mu$m and $\rho_{p}$ = 1000 kg/m$^{3}$, where we can see that the MSD exhibits a ballistic regime at short time intervals, namely, $\text{MSD} \propto \tau^{2}$ for $\tau \le t_{f}$.
At longer time intervals, the MSD asymptotically approaches a plateau value, indicating confined motions for the particles, which is due to the walls of the convection cell.
Previously, there were contrary results \cite{schumacher2008lagrangian,ni2013experimental} on pair particle dispersion in different directions because the turbulent thermal convection is anisotropic with vertically rising or falling plumes.
Here, we further examine whether the group of particles dispersion properties is isotropic.
We decompose the distance vector $\mathbf{r}$ into a lateral ($\mathbf{r}_{x}$) and vertical ($\mathbf{r}_{y}$) part and calculate the MSD in the lateral and vertical directions separately as
\begin{equation}
\text{MSD}_{x}(\tau)=\langle [\mathbf{r}_{x}(t+\tau)-\mathbf{r}_{x}(t)]^{2} \rangle, \ \ \ \ \
\text{MSD}_{y}(\tau)=\langle [\mathbf{r}_{y}(t+\tau)-\mathbf{r}_{y}(t)]^{2} \rangle
\end{equation}
From Fig. \ref{MSD}(b), we can see that the MSD is isotropic at short time intervals, while the differences between $\text{MSD}_{x}$ and $\text{MSD}_{y}$ are apparent at longer time intervals.
We can also roughly estimate how the particle is constrained in different directions by calculating the square root of the plateau MSD value.
The results in Fig. \ref{MSD}(b) indicate that the vertical region of constraint is a bit larger than that of the lateral region.
The reason is that most of the suspended particles are trapped within the elliptical primary roll whose major axis has a longer vertical component than the horizontal one.
Thus, when the LSC advects the particles, they will 'travel' longer distances in the vertical direction than the lateral one.

\begin{figure}[htbp]
\centering
\includegraphics[width=13cm]{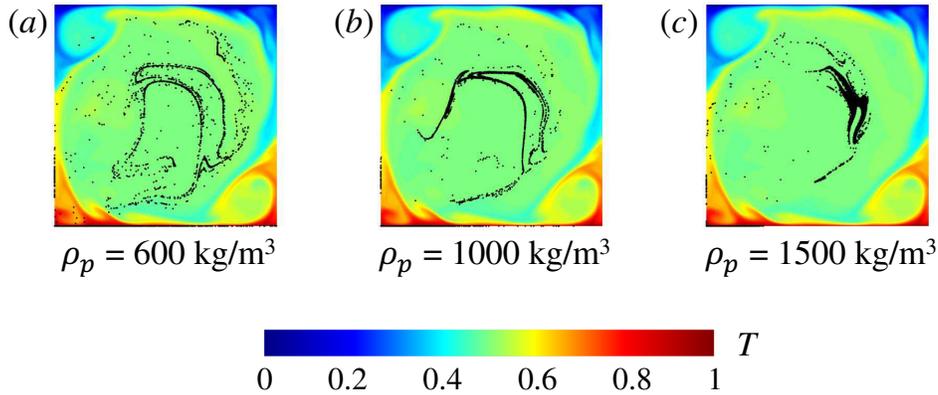}
\caption{\label{dp30um} Snapshots of the instantaneous temperature field and particles' positions ($\textit{a}$) for $\rho_{p}$ = 600 kg/m$^{3}$, ($\textit{b}$) for $\rho_{p}$ = 1000 kg/m$^{3}$, and ($\textit{c}$) for $\rho_{p}$ = 1500 kg/m$^{3}$. The diameters of these particles are 30 $\mu$m.}
\end{figure}

\begin{figure}[htbp]
\centering
\includegraphics[width=13cm]{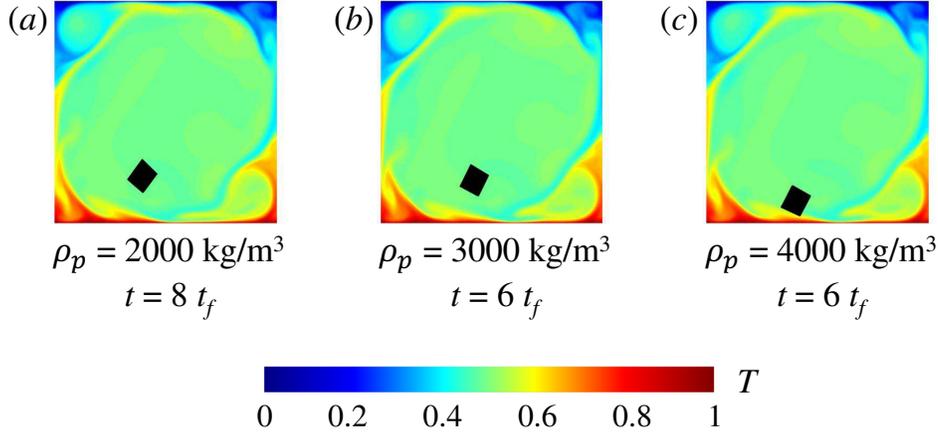}
\caption{\label{dp50um} Snapshots of the instantaneous temperature field and particles' positions ($\textit{a}$) for $\rho_{p}$ = 2000 kg/m$^{3}$ at $t = 8 \ t_{f}$, ($\textit{b}$) for $\rho_{p}$ = 3000 kg/m$^{3}$ at $t = 6 \ t_{f}$, and ($\textit{c}$) for $\rho_{p}$ = 4000 kg/m$^{3}$ at $t = 6 \ t_{f}$. The diameters of these particles are 50 $\mu$m.}
\end{figure}

The above-mentioned analysis focused on relatively small particles that will be well dispersed in the turbulence.
For relative larger particles (e.g., particles with $d_{p} = 30$ $\mu$m) in the thermal turbulence, we observe much more clear band clustering (see Fig. \ref{dp30um}).
As will be discussed in Sec. \ref{Section32}, the strong particle clustering behaviors exhibit during the transport process further results in a transition particle deposition state.
For even larger particles (e.g., particles with $d_{p} = 50$ $\mu$m), they will sediment quickly after being released in the turbulence, as shown in Fig. \ref{dp50um}.
The carrier flow minorly influences the particles' motions, and the particle group almost remains in their initial shape (namely, the square shape due to the artificial simulation setting, see Fig. 1) during the sedimentation.
Because the LSC of the convection is clockwise rotated, the deposition location of the particle group on the bottom wall will be left side offset their initial horizontal position.
We also observe that the shape of the lighter particle group will stretch more during sedimentation.
In comparison, a heavier particle group sediments faster and has a shorter horizontal offset distance for final deposition.

\subsection{\label{Section32}Particle deposition on the wall}

\begin{figure}[htbp]
\centering
\includegraphics[width=12cm]{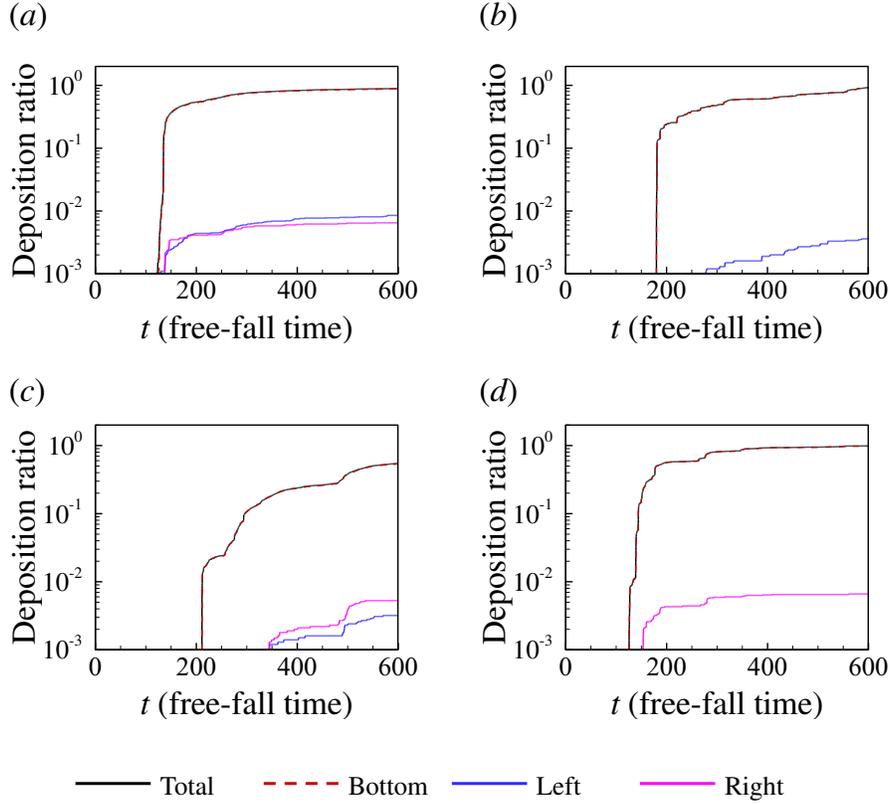}
\caption{\label{deposition_location} Time histories of particle deposition ratio for particles with $\rho_{p}$ = 1000 kg/m$^{3}$:  (\textit{a}) and (\textit{c}) $d_{p}$ = 10 $\mu$m and (\textit{b}) and (\textit{d}) $d_{p}$ = 30 $\mu$m. The convection cell is leveled in (\textit{a}) and (\textit{b}), while the cell counter-clockwise rotates 0.1$^{\circ}$ in (\textit{c}) and (\textit{d}).}
\end{figure}

We measure the particle deposition ratio as the number of deposited particles on the walls over the number of total released particles in the turbulence.
In Figs. \ref{deposition_location}(a) and \ref{deposition_location}(b), we plot the time histories of the deposition ratio for particles with $d_{p}$ = 10 $\mu$m and 30 $\mu$m, respectively.
Here, we count the number of deposited particles on the four walls of the convection cell separately, as well as their summations.
We found that most of the particles are deposited on the bottom wall, while there is no particle deposited on the top wall.
In addition, we observe a tiny portion of the particles are deposited on the left and right walls.
An interesting observation is that there are more particles deposited on the left vertical wall compared to that on the right vertical wall.
Because the LSC of the convection is clockwise rotated, the horizontal wind (from right to left) in the lower part of the convection cell will drive the particles from the right side of the cell to the left side.
When the rising hot plumes along the left vertical wall are not able to lift the particles, they will deposit on the left wall.
A similar preferential deposition pattern on hot vertical walls was also found in the side-heated convection cell \cite{puragliesi2011dns}.
To further verify the above conjecture, we measure the particle deposition ratio in a tilted convection cell, where the rotation direction of the LSC is reversed compared to that in the leveled cell.
Figures \ref{deposition_location}(c) and (d) show the particle deposition ratio in the tilted cell with vertical axis counter-clockwise rotates a small angle of 0.1$^{\circ}$ such that only the LSC rotation direction is reversed.
Still, other flows and heat transfer properties are almost not influenced by such a small tilted angle \cite{sun2005azimuthal,wang2018flow}.
In the tilted case, the hot plumes arise along the right vertical wall, and we can see that more particles are deposited on the right vertical wall.
Thus, a general conclusion is that particles prefer to be deposited on the vertical wall where the hot plumes arise.

\begin{figure}[htbp]
\centering
\includegraphics[width=12cm]{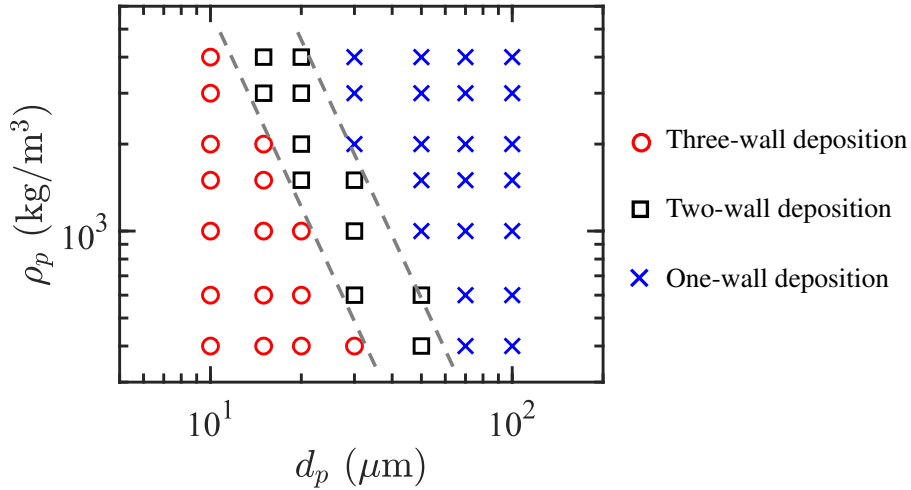}
\caption{\label{phaseDiagram} Phase diagram of the particle deposition positions on the wall. The gray dashed lines represent the rough borders between different states.}
\end{figure}

With the numerical simulations in a wide range of $d_{p}$ and $\rho_{p}$ parameter spaces, we can then obtain the phase diagram for the particle deposition positions on the walls.
As shown in Fig. \ref{phaseDiagram}, particles with smaller $d_{p}$ and $\rho_{p}$ are more easily suspended and well dispersed in the flow. Thus, the particles have chances to deposit on the left and right vertical walls, while most particles will deposit on the bottom wall due to the gravity sedimentation (denoted as 'Three-wall deposition' in the phase diagram).
For particles with larger $d_{p}$ and $\rho_{p}$, the carrier flows minorly influences them, and the particles will only deposit on the bottom wall (denoted as 'One-wall deposition' in the phase diagram).
The 'One-wall deposition' state also corresponds to the initially released particle group not well dispersed in the turbulence.
Sandwiched between the 'Three-wall deposition' and 'One-wall deposition' states is the 'Two-wall deposition' state, where particles will deposit on the bottom wall and one vertical wall at medium $d_{p}$ and $\rho_{p}$ (namely, medium $St$ and $Ar$).
This transition state of particle deposition on only one vertical wall is due to that particles exhibit cluster behavior, and they are not well dispersed in the flow compared to the cases in the 'Three-wall deposition' state.
On the other hand, in the transition state, the particles will still be majorly advected in the convection compared to the cases in the 'One-wall deposition' state, and if particles deposit, they will only deposit on vertical walls where the hot plumes arise.
For the explored parameter space of $d_{p}$ and $\rho_{p}$, we confirm that there are no particles deposited on the top wall.
From the phase diagram, we can also observe the borders between different states are strongly correlated with the $St$ and $Ar$ numbers.

\section{\label{SectionConclusions}Conclusions}

In this study, we have performed numerical simulations of particle motion in turbulent thermal convection.
Specifically, we analyzed the statistics of particle transport and deposition in 2D square RB convection cells.
The main findings are summarized as follows:

\begin{enumerate}
  \item
The suspended particles are more homogeneously distributed in the turbulence at $St$ less than $10^{-3}$, and they tend to cluster into bands for $10^{-3} \lesssim St \lesssim 10^{-2}$.
At even larger $St$, the particles' motion will be minorly influenced by the turbulence, and they will sediment quickly and deposit on the boundary walls.

  \item
At short time intervals, the MSD exhibits a ballistic regime, and it is isotropic in vertical and lateral directions.
At longer time intervals, the MSD asymptotically approaches a plateau value, indicating confined motions for the particles.
The anisotropic of MSD at longer time intervals is attributed to the tilted elliptical primary roll in which most of the particles are trapped and being advected.

  \item
We obtained a phase diagram of the particle deposition positions, and three deposition states were identified: particles deposited on three walls, two walls, and one wall.
Although most of the particles will deposit on the bottom wall, we found that there is still a tiny portion of particles deposited on the vertical wall.
Moreover, the particles preferred to deposit on the vertical wall where the hot plumes arise.
\end{enumerate}

\begin{acknowledgments}
This work was supported by the National Natural Science Foundation of China (NSFC) through Grant Nos. 11902268 and 51906044, the Fundamental Research Funds for the Central Universities of China (No. D5000200570) and the 111 project of China (No. B17037). The simulations were carried out at LvLiang Cloud Computing Center of China, and the calculations were performed on TianHe-2.
\end{acknowledgments}

\section*{Data Availability Statement}
The data that support the findings of this study are available from the corresponding author upon reasonable request.
%% and 3102019PJ002

\nocite{*}
\bibliography{myBib}% Produces the bibliography via BibTeX.

\end{document}